\documentclass[12pt]{article}

\begin{document}

\title{\bf General Formula for the Momentum Imparted to Test Particles in
Arbitrary Spacetimes}

\author{Asghar Qadir and M. Sharif\\
Department of Mathematics,\\
Quaid-i-Azam University, Islamabad,
Pakistan.}

\date{}
\maketitle
\begin{abstract}
Ehlers and Kundt have provided an approximate procedure to
demonstrate that gravitational waves impart momentum to test
particles. This was extended to cylindrical gravitational waves by
Weber and Wheeler. Here a general, exact, formula for the momentum
imparted to test particles in arbitrary spacetimes is presented.
\end{abstract}

\section{Introduction}

There has been debate whether gravitational waves really exist
[1,2]. To demonstrate the "reality" of gravitational waves, Ehlers
and Kundt [1] considered a sphere of test particles in the path of a
plane-fronted gravitational wave and showed that a constant momentum
was imparted to the test particles. A similar discussion was given
by Weber and Wheeler [2,3] for cylindrical gravitational waves. An
operational procedure embodying the same principle has been used [4]
to express the consequences of relativity in terms of the Newtonian
concept of gravitational force. The pseudo-Newtonian ($\psi$N)
"gravitational force" is defined as the vector whose intrinsic
derivative along the separation vector is the maximum tidal force,
which is given by the acceleration vector for a preferred class of
observers. The proper time integral of the force four-vector will be
the momentum four-vector. We will not, here, discuss this formalism
itself. Rather, we verify that this procedure gives the Ehlers-Kundt
result for plane-fronted gravitational waves, as may be expected on
account of the similarity of the two ideas. When we apply it to
cylindrical gravitational waves it is found that the result so
obtained is physically reasonable and gives an exact expression for
the momentum imparted to the rest particle, corresponding to the
approximation given by Weber and Wheeler.

In the free fall rest-frame the extended $\psi$N $(e\psi$N) force
four-vector is given [5] by
\begin{equation}
F_0=m[(\ln Af)_{,0}+g^{ik}g_{jk,0}g^{jl}g_{il,0}/4A],\quad F_i=m(\ln
\sqrt{g_{00}})_{,i},
\end{equation}
where $A=(\ln\sqrt{-g})_{,0}$ and $g=det(g_{ij})$. Thus the momentum
four-vector, $p_{\mu}$, is
\begin{equation}
p_{\mu}=\int F_{\mu}dt.
\end{equation}

\section{Plane-Fronted Waves}

The metric for plane-fronted gravitational waves is [6,7]
\begin{equation}
ds^2=dt^2-dx^2-L^2(t,x)\{exp[2\beta(t,x)]dy^2+exp[-2\beta(t,x)]dz^2\},
\end{equation}
where $L$ and $\beta$ are arbitrary functions subject to the vacuum
Einstein equations,
\begin{equation}
L_{,\alpha\alpha}+L\beta^2_{,\alpha}=0,~~~\alpha=0,1.
\end{equation}
Since $L$ and $\beta$ are functions of $u=t-x$, eqs.(4) reduce to
the single euqation
\begin{equation}
L_{uu}+L\beta^2_{\mu}=0
\end{equation}
From eqs.(1) and (5)
\begin{equation}
F_{0}=-m(\ddot{L}+\dot{\beta^2L})/\dot{L}=0,\quad F_{1}=0,
\end{equation}
where a dot denotes differentiation with respect to $t$.
Consequently the momentum four-vector becomes $p_{\mu}= constant$.
Thus there is a constant energy and momentum. The constant, here,
determines the "strength" of the wave. This exactly coincides with
the Ehlers-Kundt method in which they demonstrate that the test
particles acquire a constant momentum and hence a constant energy,
from the plane gravitational wave.

\section{Cylindrical Gravitational Waves}

The metric for cylinder gravitational waves is [2,3]
\begin{equation}
ds^2=exp[2(\gamma-\psi)]dt^2-exp[2(\gamma-\psi)]d\rho^2-\rho^2
exp(-2\psi)d\phi^2-exp(2\psi)dz^2,
\end{equation}
where $\gamma$ and $\psi$ are arbitrary functions of the time and
radial coordinates, $t$ and $\rho$, subject to the vacuum Einstein
equations,
\begin{equation}
\psi''+\frac{1}{\rho}\psi'-\ddot{\psi}=0,\quad \gamma'=\rho(\psi'^2+
\dot{\psi^2}),\quad\dot{\gamma}=2\rho\dot{\psi}\psi'.
\end{equation}
where a dot denotes differentiation with respect to time and a
prime differentiation with respect to $\rho$.

The solution of eqs.(8) is given by [3]
\begin{eqnarray}
\psi&=&AJ_0(\omega \rho)\cos(\omega t)+ BY_{0}(\omega \rho)\sin(\omega t), \nonumber\\
\gamma&=&\frac{1}{2} \omega\rho\{(A^2 J_0
J'_0-B^2Y_0Y'_0)\cos(2\omega t)-AB [(J_0Y'_0+Y_0J'_0) \sin(2\omega
t)\nonumber\\
&-&2(J_0Y'_0-Y_0J'_0)\omega t]\},
\end{eqnarray}
where $A$ and $B$ are arbitrary constants corresponding to the
strength of the cylindrical gravitational waves, $J_{0}$ and $Y_{0}$
are the Bessel function and the Neumann function of zero order
respectively. Here a prime denotes differentiation with respect to
$\omega\rho,~\omega$ being the angular frequency.

From eqs.(1)
\begin{eqnarray}
F_0&=&-m\{\omega[AJ_0\cos(\omega t)+BY_0 \sin(\omega t)]-2\rho
\omega[(A^2J_0J'_0 - B^2Y_0Y'_0)\cos(2 \omega t) \nonumber\\
&-&AB(J_0Y'_0+Y_0J'_0)\sin(2 \omega t)]+ 2[AJ_{0}\sin(\omega
t)\nonumber\\
&-& B Y_0 \cos(\omega t)]^2/AJ_0 \sin(\omega t)-B Y_0 \cos(\omega
t)\nonumber\\
&+&2\omega\rho[AJ_{0}\sin(\omega t)-B Y_0\cos(\omega t)][AJ'_0
\cos(\omega t)+BY'_0 \sin(\omega t)]\},\\
F_1&=& m \{AJ'_{0}\cos(\omega t)+BY'_{0}\sin(\omega t) -
\frac{1}{2}[(A^2J_0J'_0-B^2Y_0Y'_0)\nonumber\\
&+&\omega\rho(A^2J_0J'_0-B^2Y_0Y'_0)']
\cos(2\omega t), \nonumber\\
&-& \frac{1}{2}AB[2(J_0 Y'_0+Y_0J'_0)-\omega
\rho(J_0Y'_0+Y_0J'_0)']
\sin(2\omega t) \nonumber\\
&-&\frac{1}{2}AB[4(J_0Y'_0-Y_0J'_0)+2\omega \rho
(J_0Y'_0-Y_0J'_0)']\omega t\}.
\end{eqnarray}
The corresponding $p_0$ and momentum, $p_1$, imparted to the test
particle is
\begin{eqnarray}
p_0&=&-m[\ln AJ_0 \sin(\omega t)-BY_0\cos(\omega
t)\mid+\ln|1-2\omega
\rho[AJ_0'\cos(\omega t)\nonumber\\
&+&BY'_0\sin(\omega t)]|(1+\frac{A^2J_0J_0'+B^2Y_0Y_0'}{\omega
\rho(A J'^2_0+BY'^2_0)})\nonumber\\
&-&\frac{AB(AJ_0Y_0Y'_0+Y_0J'_0-A\omega\rho
J_0J'_0Y'_0-J_0Y'_0)}{\omega\rho(AJ'^2_0
+BY'^2_0)\sqrt{1-4\omega^2\rho^2(A^2J'^2_0+B^2Y'^2_0)}}\nonumber\\
&\times&\tan^{-1}|\frac{(1+2A\omega\rho J'_0)\tan(\frac{1}{2}\omega
t)-2B\omega\rho Y'_0}
{\sqrt{1-4\omega^2\rho^2(A^2J'^2_0+B^2Y'^2_0)}}|\nonumber\\
&+&\frac{AB(J'_0Y_0-J_0Y'_0)}{\rho
(A^2J_0^{'2}+B^2Y^{'2}_{0})}+f_1(\omega\rho)],\\
p_1&=&\frac{m}{\omega}\{[AJ'_{0}\sin(\omega t)-BY'_0\cos(\omega
t)]-\frac{1}{4}[A^2J_0J'_0-B^2Y_0Y'_0)\nonumber\\
&+&\rho\omega(A^2J_0J'_0-B^2Y_0Y'_0)']\sin(2 \omega t)
-\frac{1}{2}AB[J_0Y'_0+Y_0J'_0)\nonumber\\
&+&\rho\omega(J_0Y'_0+Y_0J'_0)']\cos (2 \omega t)
-AB\omega^2t^2[(J_0Y'_0-Y_0J'_0)\nonumber\\
&-&\rho\omega(J_0Y'_0-Y_0J'_0)]+f_2(\rho \omega)\},
\end{eqnarray}
where $f_1$ and $f_2$ are arbitrary constants of integration. Weber
and Wheeler [2,3] exclude solutions that contain the irregular
Bessel function, $Y_0(\omega \rho)$ as not well defined at the
origin. Taking the Weber-Wheeler solution, eqs.(12) and (13) reduce
to
\begin{eqnarray}
p_0&=&-m[\ln|AJ_0\sin(\omega t)|+(1+AK_0/\omega \rho J'_0)\ln|1-2
\omega \rho AJ'_0\cos(\omega t)|\nonumber\\
&+&f_1(\omega \rho)],\\
p_1&=&\frac{m}{\omega}\{AJ'_0\sin(\omega
t)-\frac{1}{4}A^2[J_0J'_0+\rho\omega(J_0J'_0)']\sin(2\omega t)+f_2
(\rho \omega)\}.
\end{eqnarray}
We see that the quantity $p_{1}$ given by eq.(15) can be made zero
for the small and large $\rho$ limits by choosing $f_{2}=0$. This
expression coincides with the approximate momentum expression given
by Weber and Wheeler in the large and small $\rho$ limits. Hence
this is a physically reasonable expression for the momentum imparted
to test particles by cylindrical gravitational waves. The quantity
$p_{0}$ given by eq.(14) remains finite for small $\rho$ and can
also be made finite for large $\rho$ by choosing $f_{1}=-\ln(J_0)$.
However, there is a singularity at $\omega t = n \pi$. This problem
does not arise in the general expression given by eq.(12). However,
in that case there appears a term linear in time which creates
interpretational problems. Also $ p_{0}$ and $p_{1}$ become singular
at $\rho=0$ if $B\neq0$.

\section{Conclusion}

The Ehlers-Kundt method gives the physically reasonable result that
plane gravitational waves impart a constant energy and momentum to
test particles in their path. However, it does not provide a simple
formula that can be applied to other cases. Since the essential idea
there is exactly embodied in the $e\psi$N-formalism [5] it is
reasonable to expect that it would give the same result. We have
seen that the $e\psi$N-formalism gives a physically acceptable
expression for the energy and momentum imparted to test particles by
gravitational waves which coincides with the Ehlers-Kundt approach
for the plane-fronted waves. The expression for momentum coincides
with the Weber-Wheeler [2,3] approximate result for cylindrical
waves for small and large values of $\rho$. \emph{In effect it
provides a general formula for the momentum imparted to test
particles in arbitrary spacetimes.}

The inclusion of $Y_{0}$ in the solution increases the energy at the
source and thus provides a singularity at $\rho=0$. If we do not
include the $Y_{0}$ term in the solution, $p_{0}$ becomes infinite
at various times. One would have normally taken $p_{0}$ to be the
energy of the test particle. However, that should be given by
$(m^2+p_1^2)^{1/2}$. As such the significance of $p_{0}$  is not
clear. As this is the quantity which has the problem of becoming
singular, while $p_{1}$ is well behaved we are ignoring the problem.
However, that problem does need to be resolved and an interpretation
for $p_{0}$ be provided.

There is a problem of defining energy in general relativity which is
most severe when trying to understand the energy content of
gravitational waves. As gravitational waves are solutions of the
vacuum field equations the stress-energy tensor is zero. This is the
problem regarding the reality of gravitational waves. On the one
hand it seems that gravitational waves do not carry energy but on
the other hand detectors are built to extract energy from them. If
we linearize the Einstein field equations [3,6], the complete
non-linear equations can be rewritten with the linear part on the
left side and all the non-linear terms on the right side. The
non-linear part can then be regarded as the stress-energy tensor for
the linearized wave equation [6]. Also, a $3+1$ split, such as that
of Arnowitt, Deser and Misner [8], provides an expression for the
energy density in gravitational waves. We are only providing an
expression for the energy imparted to a test particle. This
expression is based on an operational $3+1$ split. Hence this
procedure might be able to provide a handle to the problem of the
energy content in gravitational waves in particular and the
definition of energy in general. To be able to apply it successfully
the problem of $p_0$ would have to be resolved.

\vspace{.5 cm}

{\bf Acknowledgement}

\vspace{.5 cm}

We would like to thank the Pakistan Science Foundation for the
financial support during this research under the project
PSF/RES/C-QU/MATHS.(16).

\vspace{.5cm}

{\bf References}

\vspace{.5cm}

\begin{description}

\item{[1]} J. Ehlers and W. Kundt: \emph{Gravitation: An Introduction to
Current Research}, ed. L.Witten (Wiley, New York, 1962).

\item{[2]} J. Weber and J.A. Wheeler: Rev. Mod. Phys. \textbf{29}(1957)509.

\item{[3]} J. Weber: \emph{General Relativity and Gravitational Waves}
(Interscience, New York, 1961).

\item{[4]} S.M. Mahajan, A. Qadir and P.M. Valanju, Nuovo Cimento \textbf{B65}(1981)404;
A. Qadir and J. Quamar: \emph{Proc. 3rd Marcel Grossmann Meeting on
General Relativity} ed. Hu Ning (North-Holland, Amsterdam,
1983)p189;\\
J. Quamar: Ph.D. Thesis, Quaid-i-Azam University (1984).

\item{[5]} A. Qadir and M Sharif: \emph{Proc. 4th Regional Conf. on Mathematical Physics}
eds. F. Ardalan, H. Arfaei and S. Rouhani (World Scientific,
Singapore) to be published; QAUM 60/1991;\\
M. Sharif: Ph.D. Thesis, Quaid-i-Azam University (1991).

\item{[6]} C.W. Misner, K.S. Thorne and J.A. Wheeler: \emph{Gravitation}
(Freeman, San Francisco, 1973).

\item{[7]} D. Kramer, H. Stephani, E. Herlt and M. MacCallum: \emph{Exact Solutions
of Einstein's Field Equation} (Cambridge Univ. Press, Cambridge,
1979).

\item{[8]} R. Arnowitt, S. Deser and C.W. Misner: Phys. Rev.
\textbf{117}(1960)1595; ibid \textbf{118}(1960)1100.
\end{description}
\end{document}